\documentclass[twocolumn,letter]{jpsj2} %% two-column for letters
\def\runauthor{H. Fukazawa {\it et al.}}

\title{%
$^{75}$As NMR study of the ternary iron arsenide BaFe$_{2}$As$_{2}$
}

\author{%
Hideto~Fukazawa$^{1}$\thanks{E-mail address: hideto@nmr.s.chiba-u.ac.jp}, 
Kenji~Hirayama$^{1}$,
Kenji~Kondo$^{1}$,
Takehiro~Yamazaki$^{1}$,
Yoh~Kohori$^{1}$, \\
Nao~Takeshita$^{2}$,
Kiichi~Miyazawa$^{2,3}$,
Hijiri~Kito$^{2}$, 
Hiroshi~Eisaki$^{2}$, 
Akira~Iyo$^{2}$
}

\inst{%
$^{1}$Department of Physics, Chiba University, Chiba 263-8522, \\
$^{2}$Nanoelectronics Research Institute, National Institute of 
Advanced Industrial Science and Technology, Tsukuba 305-8562\\
$^{3}$Department of Applied Electronics, Tokyo University of Science, Noda 278-8510
}

\recdate{\today}

\abst{%
We report $^{75}$As nuclear magnetic resonance (NMR) measurements of the ternary iron arsenide BaFe$_{2}$As$_{2}$. 
$^{75}$As-NMR spectra clearly revealed that magnetic transition occurs at around 131~K in our samples, 
which corresponds to the emergence of spin density wave. 
Temperature dependence of the internal magnetic field 
suggests that the transition is likely of the first order. 
The critical-slowing-down phenomenon in the spin-lattice relaxation rate $1/T_{1}$ is not pronounced in this material. 
}

\kword{%
BaFe$_{2}$As$_{2}$, spin density wave, magnetic transition, nuclear magnetic resonance
}

\begin{document}
\maketitle

%\section{\label{sec:level1} Introduction}

The discovery of superconductivity in F-doped LaFeAsO with $T_{\rm c} = 26$~K~\cite{Kam1} 
has accelerated the world-wide investigations of the related superconductors
~\cite{Kit1,Take1,Nak1,Mat1,Ish1,Sat1,Taka1,GCh1,Ren1,Ren2,XCh1,Cru1}. 
The common feature of these compounds is the possession of FeAs layer which is analogous to CuO$_{2}$ plane 
in high superconducting transition temperature (high $T_{\rm c}$) cuprates. 
In addition, non-doped materials commonly exhibit spin density wave (SDW) or 
antiferromagnetic order with adjacent structural phase transition, 
which is also resemblance of the parent materials of high $T_{\rm c}$ cuprates.  
At present stage, the suppression of spin density wave (or antiferromagnetic order of localized moment) 
and/or the carrier control of non-doped materials 
seem to have a key role of the emergence of this new-class superconductivity. 
Recent nuclear magnetic resonance (NMR) studies of some of these superconducting oxypnictides strongly suggest that 
the superconductivity in the materials is unconventional one with nodes in the superconducting gap~\cite{Nak1,Mat1}. 

Soon after the intensive investigations of the oxypnictides, 
oxygen-free iron pnictides BaFe$_{2}$As$_{2}$~\cite{Rot1} and SrFe$_{2}$As$_{2}$~\cite{Kre1} 
were proposed as a next candidate of the parent materials of superconductors with high $T_{\rm c}$. 
The lattice parameter and the magnetic susceptibility of these materials were already reported about 25-30 years ago~\cite{Pfi1,Pfi2}. 
However, the deeper studies of the compounds have just started with consideration as the parent materials of superconductors. 
The crystal structure of these pnictides is the ThCr$_{2}$Si$_{2}$-type structure 
which is familiar in the heavy fermion systems.  
This structure possesses the similar FeAs layer to that realized in LaFeAsO. 
Moreover, both the materials exhibit the SDW anomalies 
at $T_{\rm SDW} =$~140~K (BaFe$_{2}$As$_{2}$)~\cite{Rot1,Hua1} and $T_{\rm SDW} =$~205~K (SrFe$_{2}$As$_{2}$)~\cite{Kre1,Yan1}. 
It is important to notice that both the compounds exhibit structural phase transition 
from tetragonal one ($I4/mmm$) to orthorhombic one ($Fmmm$) simultaneously with the SDW anomaly~\cite{Rot1,Kre1,Hua1,Yan1}.  
The orders of these anomalies are reported as a second one for BaFe$_{2}$As$_{2}$~\cite{Rot1} and 
a first one for SrFe$_{2}$As$_{2}$~\cite{Kre1,Yan1}. 
However, quite recent neutron diffraction measurements of BaFe$_{2}$As$_{2}$ report 
hysteresis of tetragonal (220) peak between on-cooling sequence and on-warming sequence,
which indicates the structural order at $T_{\rm SDW}$ is of the first order also in BaFe$_{2}$As$_{2}$~\cite{Hua1}. 

The most striking feature of these compounds is that the SDW anomaly disappears and the superconductivity indeed sets in 
by hole doping, for example, K substitution for Ba~\cite{Rot2} or Sr~\cite{GCh2}. 
In order to understand the superconductivity in these doped oxygen-free iron-based pnictides, 
it is also important to study the magnetic and the electronic properties of the parent materials. 
Especially the relation between the SDW instability and the superconductivity 
should be revealed by local-probe measurements in addition to bulk measurements. 
Hence, we performed $^{75}$As-NMR measurements of BaFe$_{2}$As$_{2}$. 
%In this paper, we report $^{75}$As-NMR of BaFe$_{2}$As$_{2}$.  
$^{75}$As-NMR spectra clearly exhibited the magnetic transition at around 131~K in our samples. 
Temperature $T$ dependence of the internal magnetic field 
suggests that the transition is likely of the first order. 
The critical-slowing-down phenomenon in the spin-lattice relaxation rate $1/T_{1}$ is not pronounced in this material.

%\section{\label{sec:level2}Experimental}

The polycrystalline BaFe$_{2}$As$_{2}$ was synthesized by the high temperature and high pressure method. 
The samples were confirmed as nearly single phase by x-ray diffraction. 
The standard four-probe resistivity measurement revealed the rapid decrease of the resistivity below 131~K, 
which corresponds to the SDW anomaly~\cite{Tom1}. 
The $T_{\rm SDW}$ of our samples is slightly lower than those reported by Rotter {\it et al.}~\cite{Rot1} and Huang {\it et al.}~\cite{Hua1} 
The samples were crushed into powder for the experiments. 
The NMR experiments of the $^{75}$As nucleus ($I=3/2$, $\gamma = 7.292$~MHz/T) have been carried out 
by using phase-coherent pulsed NMR spectrometers and a superconducting magnet. 
The NMR spectra were measured both by sweeping the applied fields at a constant resonance frequency 
and by sweeping the resonance frequency at a constant applied field. 
The origin of the Knight shift $K=0$ of $^{75}$As nucleus was determined by the $^{75}$As NMR of GaAs~\cite{Bas1}. 
The $1/T_{1}$ was measured with the saturation recovery method.

%\section{\label{sec:level3}Results and Discussions} 

 \begin{figure}
  \centering
  \includegraphics[width=8cm]{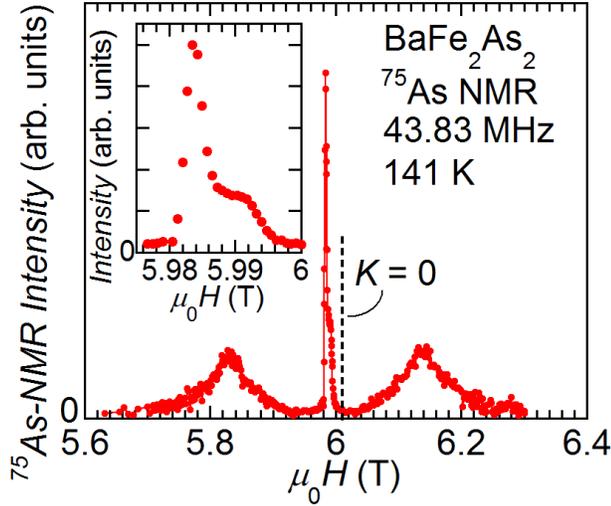}
  \caption{
  (Color online) $^{75}$As-NMR spectrum of BaFe$_{2}$As$_{2}$ at 141~K. 
  Inset shows the center line of the enlarged spectrum between 5.975 and 6~T. 
  }
 \end{figure}

In Fig.~1, we show the $^{75}$As-NMR spectrum of BaFe$_{2}$As$_{2}$ at 141~K in paramagnetic state. 
At this temperature, the crystal structure of BaFe$_{2}$As$_{2}$ is tetragonal~\cite{Rot1,Hua1}. 
The obtained spectrum is a powder pattern expected for weak electric quadrupole coupling~\cite{Car1}. 
The nuclear-spin Hamiltonian with external magnetic field $H_{\rm ext}$ is given by   
$\mathcal{H} = -\gamma H_{\rm ext}I_{z} + \frac{e^{2}qQ}{4hI(2I-1)}\left( 3I_{z'}^{2}-I(I+1)\right), $
%\begin{equation}
% \mathcal{H} = -\gamma H_{\rm ext}I_{z} + \frac{e^{2}qQ}{4hI(2I-1)}\left( 3I_{z'}^{2}-I(I+1)\right) \nonumber
%\end{equation}
where $h$,$eq$, $eQ$ represent the Planck constant, the electric field gradient (EFG) and 
the nuclear quadrupole moment, respectively.
The principal axis of the EFG is along the crystal $c$ axis since As site above $T_{\rm SDW}$ 
has a local four-fold symmetry around the $c$ axis~\cite{Rot1,Hua1}. 
Because of the same reason, the $\mathcal{H}$ above $T_{\rm SDW}$ does not contain the asymmetry parameter $\eta$ of EFG. 
Sharp central peak was observed at around 5.98~T. 
In addition, rather broader satellite peaks were observed at around 5.83 and 6.13~T, 
which is due to the first-order perturbation effect of the electric quadrupole term against the Zeeman term in $\mathcal{H}$. 
%and suggests that the crystals are not fully oriented along the magnetic easy axis. 
From the difference of the resonance fields of the satellite lines, a nuclear quadrupole resonance frequency,
$\nu_{\rm Q} \equiv \frac{3e^{2}qQ}{2hI(2I-1)}$, was estimated as 2.2~MHz. 
This value is about 4-5 times smaller than the values in iron-based oxypnictide superconductors~\cite{Nak1,Mat1,Gra1,Muk1}. 
This smaller value  might suggest the low career density in the parent materials. 
Similar discussion was done in the nuclear quadrupole resonance 
of the oxygen-deficient iron oxypnictides~\cite{Muk1} and the high $T_{\rm c}$ cuprates~\cite{Zhe1}. 

In the inset of Fig.~1, we show the center line of the enlarged spectrum between 5.975 and 6~T.
This line shape is due to the second-order perturbation effect of the electric quadrupole term. 
The large peak at around 5.984~T and the edge-like structure at around 5.993~T originate from 
the resonance components perpendicular and 41.8$^{\rm o}$-inclined to the principal axis (crystal $c$ axis) of the EFG, respectively. 
The enhancement of the perpendicular component, which is parallel to the crystal $ab$ plane, 
is due to the partial orientation of the crystals in the $H_{\rm ext}$. 
This suggests that the magnetic easy axis of BaFe$_{2}$As$_{2}$ is within the $ab$ plane, 
which is consistent with the magnetic susceptibility parallel to the $ab$ plane is 
about 1.5 times larger than that parallel to the $c$ axis~\cite{Wan1}. 

 \begin{figure}
  \centering
  \includegraphics[width=8cm]{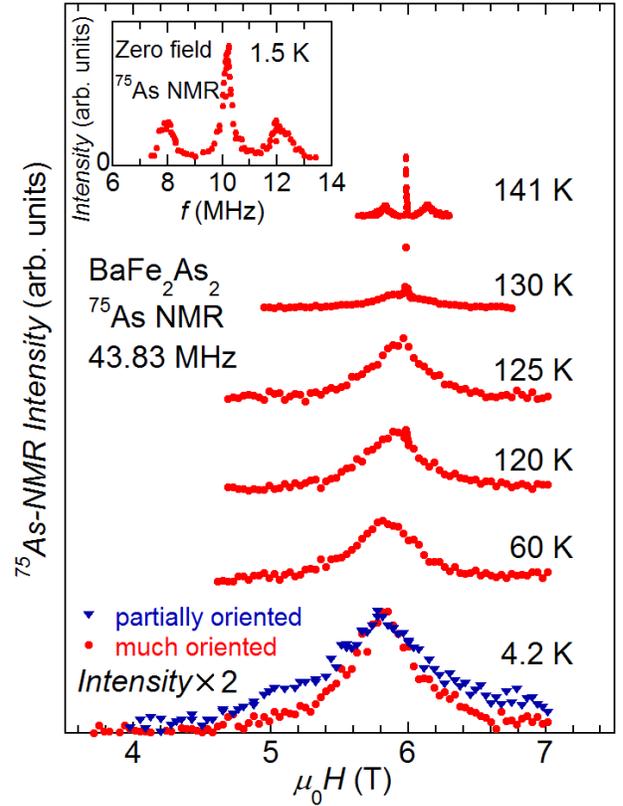}
  \caption{
  (Color online) $^{75}$As-NMR spectra of BaFe$_{2}$As$_{2}$ at various temperatures. 
  Circles denote the spectrum with much oriented powders in external magnetic field 
  and triangles the spectrum with partially oriented powders. 
  Inset shows the zero-external-field $^{75}$As-NMR spectrum at 1.5~K.
  }
 \end{figure}

In Fig.~2, we show the $^{75}$As-NMR spectra of BaFe$_{2}$As$_{2}$ at various temperatures. 
The spectrum broadens below 130~K.  
This is clearly ascribable to the SDW magnetic ordering below $T_{\rm SDW} = 131$~K. 
The quite weak signals from the center line observed in the paramagnetic state remains at around 6~T even in the magnetically ordered state. 
Since the signal intensity of this center peak below $T_{\rm SDW}$ becomes about 1/10 of that above $T_{\rm SDW}$, 
we may speculate that the SDW ordered state and the paramagnetic state separately exist just below $T_{\rm SDW}$, 
which implies that the order of the SDW anomaly in BaFe$_{2}$As$_{2}$ is of the first order. 
Note that the spectral broadening below $T_{\rm SDW}$ cannot be explained 
by the change of the $\nu_{\rm Q}$ associated with the structural phase transition at $T_{\rm SDW}$. 
Because the $\nu_{\rm Q}$ is roughly inversely proportional to unit cell volume 
and the change of the unit cell volume associated with the SDW anomaly is at most 5\%, 
the change of the $\nu_{\rm Q}$ gives the spectral broadening of at most 0.02~T at half maximum of the intensity at 6~T. 
This is clearly much less than the experimental spectral broadening of 0.5~T at 120~K. 

In Fig.~2, we also show two $^{75}$As-NMR spectra at 4.2~K. 
One is obtained with much oriented powders and another is with partially oriented powders. 
The orientation of the micro crystals in powders was estimated from the fraction of the perpendicular component 
of the center line spectrum in the paramagnetic state. 
Clear difference of the spectra was observed. 
The broader spectrum with partially oriented powders indicates 
the larger projection of the internal magnetic field $H_{\rm int}$  at As site along the $H_{\rm ext}$. 
Since the $c$ axis of each micro crystal in fully oriented powders is perpendcicular to the $H_{\rm ext}$, 
we deduce that the $H_{\rm int}$ at As site directs the crystal $c$ axis. 
Recent neutron diffraction measurements revealed the ordered moment of 0.87$\mu_{\rm B}$ (Bohr magneton) at Fe site 
with the $q$ vector of $(1,0,1)$ for the orthorhombic structure~\cite{Hua1}. 
Indeed, the antiferromagnetic Fe ordered moment within the $ab$ plane will make the $H_{\rm int}$ at As site parallel to the $c$ axis 
because the As site locates at the top of the AsFe$_{4}$ pyramid. 
The sharper $^{75}$As-NMR spectrum at 4.2~K is obtained in a condition that 
the $H_{\rm ext}$ and the $H_{\rm int}$ orient nearly perpendicular. 
Hence, the line width of the spectrum in Fig.~2 arises from the small angle distribution of the alignment, 
which is reduced from the bare $H_{\rm int}$. 
However, the $T$ dependence of the line width corresponds to that of the $H_{\rm int}$. 
%Since the fully oriented powders are parallel to the crystal $ab$ plane, 
%we deduce that the $H_{\rm int}$ at As site directs the crystal $c$ axis. 
%This implies that the ordered moment at Fe site is within the $ab$ plane 
%because the As site locates at the top of the AsFe$_{4}$ pyramid. 
%This is consistent with the fact that the magnetic easy axis of this compound is within $ab$ plane. 
%Recent neutron diffraction measurements revealed the ordered moment of 0.87$\mu_{\rm B}$ (Bohr magneton) at Fe site 
%with the $q$ vector of $(1,0,1)$ for the orthorhombic structure~\cite{Hua1}. 
%This is consistent with the fact that the magnetic easy axis of this compound is within $ab$ plane. 
Assuming this ordered structure and the moment along the $a$ or the $b$ axis, 
we evaluated that the dipole field at As site is approximately 0.3~T. 
This is less than the $H_{\rm int}$ estimated from the spectral width of the $^{75}$As-NMR. 
The contribution of conduction electrons is involved in the actual $H_{\rm int}$, 
though the direction of the $H_{\rm int}$ depends on that of the ordered moments and the ordered structure. 

In the inset of Fig.~2, we show zero-field $^{75}$As-NMR spectrum at 1.5~K. 
These narrow center line and broad satellites indicate that the $H_{\rm int}$ at As site had a magnitude of about 1.3~T 
and its orientation is the maximum EFG direction at As site ($c$ axis). 
The narrow lines suggest that the magnetically ordered state is basically formed with the commensurate $q$ vector, 
which is consistent with neutron scattering measurements~\cite{Hua1}.
The origin of the broader satellites is probably due to the slight distribution of the EFG in the orthorhombic structure 
through the first order structural phase transition. 
Note that the broadening of the NMR spectra in the $H_{\rm ext}$ is due to the distribution of 
the projection component of the $H_{\rm int}$ along the $H_{\rm ext}$. 
Quite recent $^{75}$As-NMR measurements of single crystals and polycrystals of BaFe$_{2}$As$_{2}$ by Baek {\it et al.} 
apparently revealed that the magnetically ordered state is incommensurate~\cite{Bae1}. 
However, our zero-field $^{75}$As-NMR spectrum indicates that the magnetically ordered state is commensurate. 

%In the inset of Fig.~2, the temperature $T$ dependence of the spectral intensity of the center peak 
%observed in the paramagnetic state at around 5.984~T. 
%The data was obtained on warming and on cooling with the rate of 5~K/min. 
%The signal intensity drastically decreases below about 135~K. 
%The middle point of the transition is 131~K and this temperature is 
%consistent with $T_{\rm SDW}$ determined by resistivity in our samples~\cite{Tom1}. 
%No clear hysteresis of $T$ dependence of the signal intensity was observed. 
%This experimental result suggests that the order of the SDW anomaly in BaFe$_{2}$As$_{2}$ is of the second order.
%Since our temperature control rate is three times slower than that operated by Huang {\it et al.}~\cite{Hua1}, 
%the apparent hysteresis observed in experiments by Huang {\it et al.}~\cite{Hua1} is possibly due to the rapid temperature control rate. 

 \begin{figure}
  \centering
  \includegraphics[width=8cm]{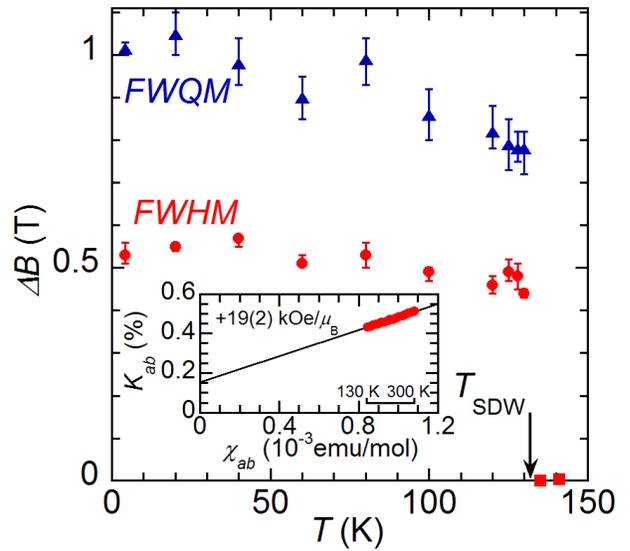}
  \caption{
  (Color online) $T$ dependence of the full width at half maximum $FWHM$ (closed circle) and 
  quarter maximum $FWQM$ (closed triangle) of the $^{75}$As-NMR spectra of BaFe$_{2}$As$_{2}$. 
  Closed squares denote the $FWHM$ of the center line in the paramagnetic state. 
  Inset shows Knight shift $K_{ab}$ along the $ab$ plane of $^{75}$As-NMR of BaFe$_{2}$As$_{2}$ 
  versus the magnetic susceptibility $\chi_{ab}$~\cite{Wan1}. 
  }
 \end{figure}

In Fig.~3, we show the $T$ dependence of the full width at half maximum $FWHM$ and quarter maximum $FWQM$ 
of the $^{75}$As-NMR spectra with much oriented powders. 
As already described above, spectral width can be considered as the rough measure of 
the distribution of the internal magnetic field $H_{\rm int}$ at As site. 
The $FWHM$ are nearly $T$ independent and the $FWQM$ slightly decreases on warming between 4.2 and 130~K. 
Then they abruptly decrease from finite field toward zero field at around $T_{\rm SDW}$. 
This discontinuous tendency of the magnetic internal field at $T_{\rm SDW}$ 
can be interpreted that the order of the SDW anomaly is likely of the first one.

% \begin{figure}
%  \centering
%  \includegraphics[width=8cm]{Fig4_BaFe2As2_Kab_chiab_5o98T.eps}
%  \caption{
%  (Color online) 
%  }
% \end{figure}

In the inset of Fig.~3, we show the Knight shift $K_{ab}$ along the $ab$ plane of $^{75}$As-NMR of BaFe$_{2}$As$_{2}$ 
versus $\chi_{ab}$, magnetic susceptibility parallel to the $ab$ plane above $T_{\rm SDW}$. 
The $K_{ab}$ above $T_{\rm SDW}$ was obtained from the perpendicular component of 
the center line of the frequency-swept spectra at a constant applied field of 5.98~T.  
The second order quadrupole effect was taken into account to obtain $K_{ab}$~\cite{Car1}. 
The data of $\chi_{ab}$ was taken from ref.~\ref{Wang}. 
The hyperfine coupling constant $A_{ab}$ parallel to the $ab$ plane was evaluated from the Knight shift at As site parallel to the $ab$ plane 
and the magnetic susceptibility parallel to the $ab$ plane by assuming the following formulae, 
$K_{ab}(T) = \frac{A_{ab}^{d}}{N_{\rm A}\mu_{\rm B}}\chi_{ab}^{d}(T)
+ \frac{A_{ab0}}{N_{\rm A}\mu_{\rm B}}\chi_{ab0}$ 
$\left( \chi_{ab}(T) = \chi_{ab}^{d}(T) + \chi_{ab0}\right).$   
%$$K_{ab}(T) = \frac{A_{ab}^{d}}{N_{\rm A}\mu_{\rm B}}\chi_{ab}^{d}(T)
%+ \frac{A_{ab0}}{N_{\rm A}\mu_{\rm B}}\chi_{ab0}, $$    
%$$\chi_{ab}(T) = \chi_{ab}^{d}(T) + \chi_{ab0}.$$    
Here, $N_{\rm A}$ represents Avogadro number. 
The evaluated hyperfine coupling constant $A_{ab}^{d}$ originating from 
the coupling between $^{75}$As nuclear spin and 3$d$ conduction electron of Fe is +19(2)~kOe/$\mu_{\rm B}$. 
This is the same magnitude of the coupling constant as that reported for the $^{75}$As-NMR of LaFeAsO$_{0.9}$F$_{0.1}$~\cite{Gra1}. 
%Note that the estimation of $A_{ab}$ has an ambiguity 
%since the $T_{\rm SDW}$ of our samples are slightly different from that of single crystals grown by the authors of ref.~\ref{Wang}. 
%However, the reasonable magnitude of the $A_{ab}^{d}$ supports that the estimation is roughly adequate. 

 \begin{figure}
  \centering
  \includegraphics[width=8cm]{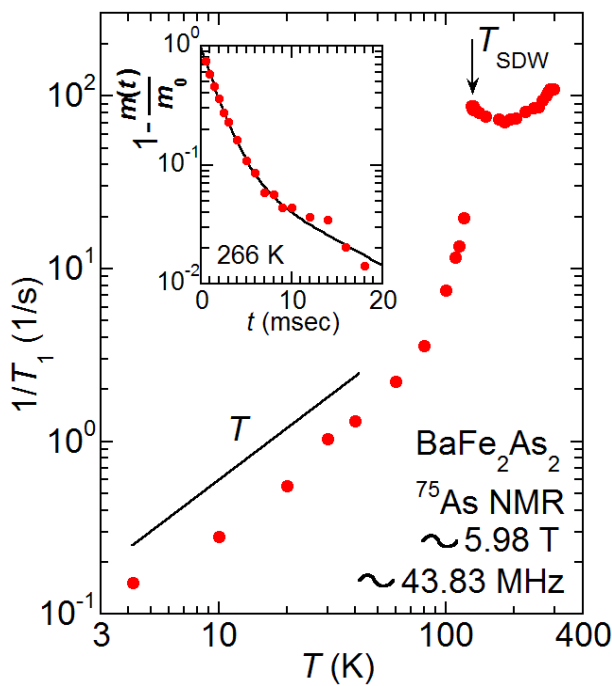}
  \caption{
  (Color online) $T$ dependence of spin-lattice relaxation rate $1/T_{1}$ of $^{75}$As of BaFe$_{2}$As$_{2}$. 
  The inset shows the typical recovery curve obtained at 266~K. 
  }
 \end{figure}

In Fig.~4, we show $T$ dependence of $1/T_{1}$ of $^{75}$As of BaFe$_{2}$As$_{2}$.
We obtained $T_{1}$ at a fixed frequency of 43.83~MHz with the external field of 5.75-5.98~T in the $T$ range of 4.2-131~K and 
at a fixed field of 5.98~T with the frequency of 43.83-43.87~T in the $T$ range of 131-300~K. 
The nuclear magnetization recovery curve was fitted by a following double exponential function 
as expected for the center line of the spectrum of the nuclear spin $I=3/2$ of the $^{75}$As nucleus~\cite{Sim1}, 
%$$1-\frac{M(t)}{M_{0}} = 0.1\exp\left( -\frac{t}{T_{1}}\right)  + 0.9\exp\left( -\frac{6t}{T_{1}}\right) ,$$
$1-\frac{m(t)}{m_{0}} = 0.1\exp\left( -\frac{t}{T_{1}}\right)  + 0.9\exp\left( -\frac{6t}{T_{1}}\right) ,$
where $m(t)$ and $m_{0}$ are nuclear magnetizations after time $t$ and enough time from the NMR saturation pulse. 
In the inset of Fig.~4, we show typical recovery curve obtained at 266~K. 
Clearly the data are well fitted to the above ideal curve with a single $T_{1}$ component.  
We used this formula even below $T_{\rm SDW}$, where the recovery curve should be more complicated and the concrete form cannot be derived. 
However, the analysis below $T_{\rm SDW}$ yielded good fitting with a single $T_{1}$ component. 
$1/T_{1}$ exhibits rapid decrease between 180 and 300~K. 
The slope of the $1/T_{1}$ is greater than $T$ linear dependence. 
This anomalous $T$ dependence is quite different from that reported in the related parent compound LaFeAsO
where $1/T_{1}$ of $^{139}$La is nearly $T$ independent sufficiently above $T_{\rm SDW}$~\cite{Nak1}. 
Below about 180~K, $1/T_{1}$ exhibits gradual increase and sudden decrease below 131~K~$=T_{\rm SDW}$. 
%This $T$ dependence around $T_{\rm SDW}$ is ascribable to a critical-slowing-down phenomenon associated with the SDW magnetic order. 
Note that the critical-slowing-down phenomenon in this material is not pronounced, 
which is contrast to the case of LaFeAsO in which the phenomenon was more clearly observed in $1/T_{1}$ of $^{139}$La~\cite{Nak1}.   
The steep decrease of $1/T_{1}$ below $T_{\rm SDW}$ is probably due to a gap formation of the SDW at part of the Fermi surface. 
%However, it should be noted that the comparison of $1/T_{1}$ between $T < T_{\rm SDW}$ and $T > T_{\rm SDW}$ is quite difficult 
%since we used different part of the spectrum to evaluate $1/T_{1}$ below $T_{\rm SDW}$ and above $T_{\rm SDW}$.
%We analyzed the data between 100~K and $T_{\rm SDW}$ by assuming the gap function $1/T_{1} \propto \exp(-\Delta/T)$. 
%This analysis yielded $\Delta \simeq 1000$~K which is much larger than the expected SDW gap of about 200~K. 
$1/T_{1}$ below about 100~K is nearly proportional to $T$. 
This is attributable to the relaxation due to the remaining conduction electron at the Fermi level even below $T_{\rm SDW}$. 

In BaFe$_{2}$As$_{2}$ and its isomorph SrFe$_{2}$As$_{2}$, the magnetic anomaly and the structural anomaly coincide 
with each other at $T_{\rm SDW}$~\cite{Rot1,Kre1,Hua1,Yan1}. 
This is in contrast to the case in the related compound LaFeAsO in which the magnetic anomaly and the structural anomaly 
occur with the separation of about 10~K~\cite{Nak1}. 
Moreover, the present results of $^{75}$As NMR measurements of BaFe$_{2}$As$_{2}$ support that 
the magnetic transition at $T_{\rm SDW}$ is likely of the first order. 
Further investigation of the relation between the coincidence/separation of the magnetic anomaly and the structural anomaly 
and the order of the phase transition in the parent materials is quite important to understand the superconductivity 
adjacent to the magnetic and the structural anomalies in the iron-based oxypnictides/pnictides.  

In summary, we performed $^{75}$As NMR measurements of the ternary iron arsenide BaFe$_{2}$As$_{2}$ 
which is a parent compound of new-class iron-based superconductors. 
$^{75}$As-NMR spectra clearly revealed that magnetic transition occurs at around 131~K in our samples, 
which corresponds to the emergence of SDW. 
$T$ dependence of the $H_{\rm int}$ suggests that the transition is likely of the first order. 
However, it is still an open question whether this transition is of the first order or of the second one. 
The detailed specific heat measurements will answer this question and is strongly required. 
The critical-slowing-down phenomenon in $1/T_{1}$ is not pronounced in this compound. 
Finally we comment about the pressure effect on these iron-based oxypnictides/pnictides. 
Not only doping study but also recent pressure study of the oxypnictides revealed that 
the pressure much affects these materials in the electronic states in the broad pressure range up to about 30~GPa~\cite{Take1,Taka1}. 
This indicates that the subsequent NMR/NQR studies of iron-based oxypnictides/pnictides under high pressure 
requires such high pressure of 10~GPa class. 
Recent development of high pressure NMR/NQR technique by the authors' group~\cite{Fuk1,Hir1} will 
help for giving further insights into the understanding of these fascinating iron-based oxypnictides/pnictides.

%\section{Summary}

%\section*{Acknowledgment}

This work is supported by a Grant-in-Aid for Scientific Research from the MEXT.

\end{document}